\documentclass[aps,prl,twocolumn,showpacs,superscriptaddress]{revtex4-1}
\usepackage[latin1]{inputenc}
\usepackage[english]{babel}
\usepackage{bm,graphicx,amsmath,amssymb,color,epstopdf}

\begin{document}

\title{Too Close to Integrable: Crossover from Normal to Anomalous Heat Diffusion} 
\author{Stefano Lepri}
\affiliation{Consiglio Nazionale delle Ricerche, Istituto dei Sistemi Complessi, Via Madonna del Piano 10 I-50019 Sesto Fiorentino, Italy} 
\affiliation{Istituto Nazionale di Fisica Nucleare, Sezione di Firenze, via G. Sansone 1 I-50019, Sesto Fiorentino, Italy}
\author{Roberto Livi}
\affiliation{ Dipartimento di Fisica e Astronomia and CSDC, Universit\`a di Firenze, via G. Sansone 1 I-50019, Sesto Fiorentino, Italy}
\affiliation{Consiglio Nazionale delle Ricerche, Istituto dei Sistemi Complessi, Via Madonna del Piano 10 I-50019 Sesto Fiorentino, Italy} 
\affiliation{Istituto Nazionale di Fisica Nucleare, Sezione di Firenze, via G. Sansone 1 I-50019, Sesto Fiorentino, Italy}
\author{Antonio Politi}
\affiliation{Institute for Complex Systems and Mathematical Biology \& SUPA
University of Aberdeen, Aberdeen AB24 3UE, United Kingdom}

\date{\today}

\begin{abstract}
Energy transport in one-dimensional chains of particles 
with three conservation laws is generically anomalous and 
belongs to the Kardar-Parisi-Zhang dynamical universality
class. Surprisingly, some examples where an apparent normal heat 
diffusion is found over a large range of length scales were
reported. We propose a novel physical explanation of these intriguing observations. 
We develop a scaling analysis which explains how this may happen in
the vicinity of an integrable limit, such as, but not only, the famous Toda model.
In this limit, heat transport is mostly supplied by quasi-particles
with a very large mean free path $\ell$. 
Upon increasing the system size $L$, three different regimes can be observed: 
a ballistic one, an intermediate diffusive range, and, eventually, the  
crossover to the anomalous (hydrodynamic) regime.        
Our theoretical considerations are supported by numerical simulations of
a gas of diatomic hard-point particles for almost equal masses 
and of a weakly perturbed Toda chain. Finally, we discuss the 
case of the perturbed harmonic chain, which exhibits a yet different scenario.
\end{abstract}
\maketitle

After more than twenty years of theoretical research, there is a general consensus 
that energy transport in one and two-dimensional systems is anomalous, meaning that 
Fourier's law is invalid \cite{LLP03,DHARREV,Lepri2016}. Numerics \cite{Lepri03,Wang2011,Liu2014}
as well as hydrodynamic \cite{Narayan02,Delfini06} and kinetic \cite{Pereverzev2003,Nickel07,Lukkarinen2008} theories
consistently indicate that the nonlinear interactions of fluctuations
of conserved quantities yield, in reduced space-dimension, non standard relaxation and transport properties, 
even in the linear response regime. 
The main signature of the anomaly is the divergence of the thermal conductivity $\kappa$ with the system size $L$,
i.e. a superdiffusive heat transport.
In one-dimension, although this is a genuine many-body problem, it can be described effectively  
as an ensemble of L\'evy particles, namely random walkers performing 
free ballistic steps with finite velocity for times that are power-law distributed \cite{Zaburdaev2015}. This description 
accounts quantitatively for several non-equilibrium 
properties, both transient and stationary \cite{Cipriani05,Lepri2011b,Dhar2013,Liu2014}.    
Remarkably, the phenomenon was shown to belong to the  
class of the famous Kardar-Parisi-Zhang (KPZ) equation \cite{VanBeijeren2012,Spohn2014}, 
suggesting a universal behavior with implications for
the theory of transport in nano-sized objects like individual 
nanowires \cite{Upadhyaya2016}, nanotubes \cite{Chang2016} or polymers \cite{Crnjar2018}. 
In this general context, nanowires
and single-walled nanotubes have been analyzed to look for deviations from the standard Fourier's
law \cite{Chang2016}. Experimental evidence of such deviations
has been reported for single-walled carbon nanotubes \cite{Chang2016, Lee2017} 
(see also Ref.~\cite{Li2017comment}). Non-trivial
length dependence of thermal conductance has been also observed in molecular chains~\cite{Meier2014}. 
Transport anomalies can be even exploited to 
achieve optimal efficiency of thermal to electric energy conversion~\cite{Casati2009,Benenti2017}.

Although the general framework is pretty well understood, there are still 
open issues that escaped so far a convincing explanation.
For definiteness, we focus on anharmonic chains, represented by a Hamiltonian of the form 
\begin{equation}
H = \sum_{n=1}^L \left[\frac{p_n^2}{2m_n} + U(q_{n+1} - q_n) \right]\, ,
\label{Hamil}
\end{equation}
where $m_n$, $q_n$ and $p_n$ are respectively the mass, displacement and momentum of the $n$th particle. For 
a generic potential $U$, this family of models should show superdiffusive heat transport 
in the KPZ universality class, as confirmed by several studies~\cite{Mendl2013,Das2014a}.
However, there is evidence of significant deviations of the dynamical exponents
in some models with hard-core potential~\cite{Hurtado2016}.
Moreover, chains allowing for bond dissociation 
(like e.g Lennard-Jones, Morse, and Coulomb potentials) unexpectedly 
display finite thermal conductivity~\cite{Savin2014,Gendelman2014}, while 
other similar potentials closely follow the prediction of anomalous scaling~\cite{Mendl2014}. 
For the double-well potential, an intermediate-energy
regime with almost diffusive transport has been reported~\cite{Xiong2016,Barik2019}.  

Another, more surprising feature is the (apparent) normal diffusive heat transport 
observed at low energies in asymmetric potentials~\cite{Zhong2012} like the 
Fermi-Pasta-Ulam-Tsingou-$\alpha\beta$ (FPUT) chain, where 
$m_n=m$ and $U_{\alpha\beta}(y)=y^2/2+\alpha y^3/3+\beta y^4/4$.
Yet, more compelling evidence of a seemingly normal transport has been found in a Toda lattice
under the action of an additional conservative noise~\cite{Iacobucci2010} and in the diatomic
Hard Point Gas (HPG)~\cite{Chen2014}  (see also \cite{Zhao2018}).
In the first context, successive studies showed that the diffusive regime is a finite-size effect, 
whereby anomalous behavior is recovered
for large enough $L$~\cite{Wang2013,Das2014} (see also the discussion
based on mode-coupling arguments in Ref.~\cite{Lee-Dadswell2015a}).
The same is demonstrated in Ref.~\cite{Miron2019} for 
a stochastic hard-core gas model.

In this Letter we present a general explanation of the counter-intuitive normal transport
observed in several models.
In a nutshell the argument runs as follows. The length-independent flux exhibited by 
integrable systems is the result of the free displacement of quasi-particles 
(the integrals of motion, such as solitons) from the hot towards the cold reservoir.
In the vicinity of the integrable limit, as a result of mutual interactions, the
quasi-particles have a finite mean free path $\ell$. Thereby, a purely
ballistic behavior can be observed only for $L<\ell$. On the other hand, $L>\ell$
is not a sufficient condition to observe a crossover towards the anomalous
behavior predicted by the above mentioned theoretical arguments.
In fact, it is necessary for $L$ to be so long that the {\it normal} flux
induced by inter-particle scattering becomes negligible.

In more quantitative terms, building upon the intuition contained in Ref.~\cite{Lepri2009}, 
we conjecture that the heat flux $J(L,\varepsilon)$ is the sum of two terms,
\begin{equation}
J(L,\varepsilon) = J_A(L,\varepsilon) + J_N(L,\varepsilon) \; ,
\label{Jtot}
\end{equation}
where $\varepsilon$ is the distance from the integrable limit, $J_A$ is the hydrodynamic contribution, 
arising from the mutual interaction among density, energy, and momentum fluctuations, and
$J_N$ is a kinetic contribution, accounting for the energy transported by the weakly interacting quasi-particles.

For $L\to\infty$, $J_A \approx L^{\eta-1}$ with $\eta=1/3$ in systems belonging to the KPZ class
\cite{VanBeijeren2012,Spohn2014}, while $\eta=1/2$ in some special cases, like
models with symmetric interaction potentials (e.g, FPUT-$\beta$ chain 
with potential $U_\beta(y)=y^2/2+y^4/4$) and models subject to conservative noise 
(e.g., the noisy harmonic \cite{BBO06,Lepri2009} or nonlinear \cite{Basile08,Iacobucci2010} chains).

On the basis of standard kinetic arguments~\cite{pitaevskii2012physical}, $J_N$ is expected to 
be a function of a single compound variable, the effective length $\xi=L/\ell$ expressed in units 
of the mean free path $\ell$, the only relevant scale in this context,
\begin{equation}
J_N(L,\varepsilon) = j_N(L/\ell) \; .
\label{JN}
\end{equation}
For $\xi\gg 1$, we expect $j_N(\xi) \propto 1/\xi$, meaning that the flux is the result of a standard diffusive process, 
while $j_N(0)$ is a finite value, meaning that the process is ballistic for system sizes smaller than the mean free
path ($\xi \ll 1$).
The entire $j_N(\xi)$ dependence is captured by the simple effective formula
\begin{equation}
j_N(\xi) = \frac{j_0}{r+ \xi}
\label{kappad}
\end{equation}
where $r$ is a constant accounting for the boundary resistance \cite{Aoki01} and $j_0$ is an  additional constant.

The vicinity to the integrable limit manifests itself as a divergence of the
mean free path, which we account for by assuming $\ell \approx \varepsilon^{-\theta}$, 
where $\theta >0$ is a system-dependent exponent.
As long as $J_A(L,\varepsilon)$ does not display any singularity for $\varepsilon \to 0$ 
(we return to this point in the final part of the Letter), we can neglect its dependence on $\varepsilon$ (for
$\varepsilon \ll 1$).
Therefore, for large $L$, Eq.~(\ref{Jtot}) can be rewritten as
\begin{equation}
J(L,\varepsilon) \;\approx \; \frac{c_A}{L^{1-\eta}} \,+\, 
 \frac{c_N}{L\varepsilon^{\theta}}  \; ,
\label{Jtot2}
\end{equation}
where $c_A$ and $c_N$ are two suitable parameters. Accordingly,
the anomalous contribution prevails only above the crossover length
$\ell_c \approx \varepsilon^{-\theta/\eta}$. For $L\le \ell_c$, heat conduction is
dominated by $j_N$. In particular, within the range $[\ell=\varepsilon^{-\theta},\ell_c]$
an apparent normal conductivity is expected, which is nothing but a finite size effect.

Now, we start the numerical analysis, focusing on models of the class
(\ref{Hamil}).
More specifically, we shall consider the HPG~\cite{Casati1986,H99,Grassberger02}, 
and the Toda chain~\cite{toda2012,Toda79}
with interaction potential $U_{T}(y)=( \mathrm{e}^{-y} + y -1 )$.

The HPG dynamics consists of successive collisions between 
neighboring particles according to the kinematic rules
\begin{eqnarray}
&&u_i'=\frac{m_i-m_{i+1}}{m_i+m_{i+1}} u_i+ \frac{2 m_{i+1}}{m_i+m_{i+1}}
u_{i+1} \quad , \label{coll} \\
&&u_{i+1}'= \frac{2 m_{i}}{m_i+m_{i+1}}
u_i-\frac{m_i-m_{i+1}}{m_i+m_{i+1}} u_{i+1}, \nonumber
\end{eqnarray}
where $u_n= \dot q_n$ and the primed variables 
	denote the values after the collision. Simulations are very efficient
since they only require keeping track of the collisions \cite{Grassberger02}. 

For equal masses ($m_n=m$), both models are 
completely integrable: in the HPG, the constants of motion are the initial velocities,  
while in the Toda model the conserved actions are given 
functions of positions and momenta \cite{Flaschka1974,Henon1974}. 
Both models can be seen as gases of 
quasi-particles: \textit{velocitons} for HPG \cite{Politi2011} 
and \textit{solitons} for Toda \cite{toda2012}. Transport is 
ballistic: $\kappa(L)$ is proportional to $L$ and
the energy-current correlation function does not decay to zero at 
large times \cite{Zotos02,Shastry2010}. Note that, according to 
the classification of 
Ref.~\cite{Spohn2018}, the two models are noninteracting and interacting, respectively.

Here below, we consider two different ways of breaking integrability
(i.e. to induce interactions among the quasi-particles):
(i) different masses, such as
a diatomic arrangement whereby $m_{n}=m_1=\frac{M}{2}(1-\delta)$ ($m_n=m_2=\frac{M}{2}(1+\delta)$) 
for odd (even) $n$~\cite{Casati1986,H99,Grassberger02};
(ii) a conservative noise through random collisions exchanging
the momenta of neighboring particles at a given rate $\gamma$ \cite{Iacobucci2010} 
(see also a related model in Ref.~\cite{Bernardin2014}). 
In the former (latter) case $\delta$ ($\gamma$) plays the role of the above mentioned closeness parameter $\varepsilon$.
In either case, only three conservation laws 
survive, momentum, energy and stretch, yielding anomalous transport and dynamical scaling
\footnote{Perturbations breaking momentum conservation 
(e.g. an external pinning potential) induce normal heat diffusion in the Toda model, although they may also display 
strong finite-size effects depending on the type of potential \cite{DiCintio2018,Dhar2019}. }.
For large enough $\delta$-values, there is 
overwhelming evidence that both the diatomic HPG and Toda~\cite{H99}
belong to the KPZ universality class.  
For the randomly perturbed Toda chain, evidence of a
diverging conductivity is solid with $\eta\approx 0.44$ for 
the large collision rate $\gamma=1$ \cite{Iacobucci2010}.

%\cite{[{Prepended text}][{appended text}]Dhar2019}

\begin{figure}[th]
%\begin{center}
\includegraphics[width=0.4\textwidth,clip=true]{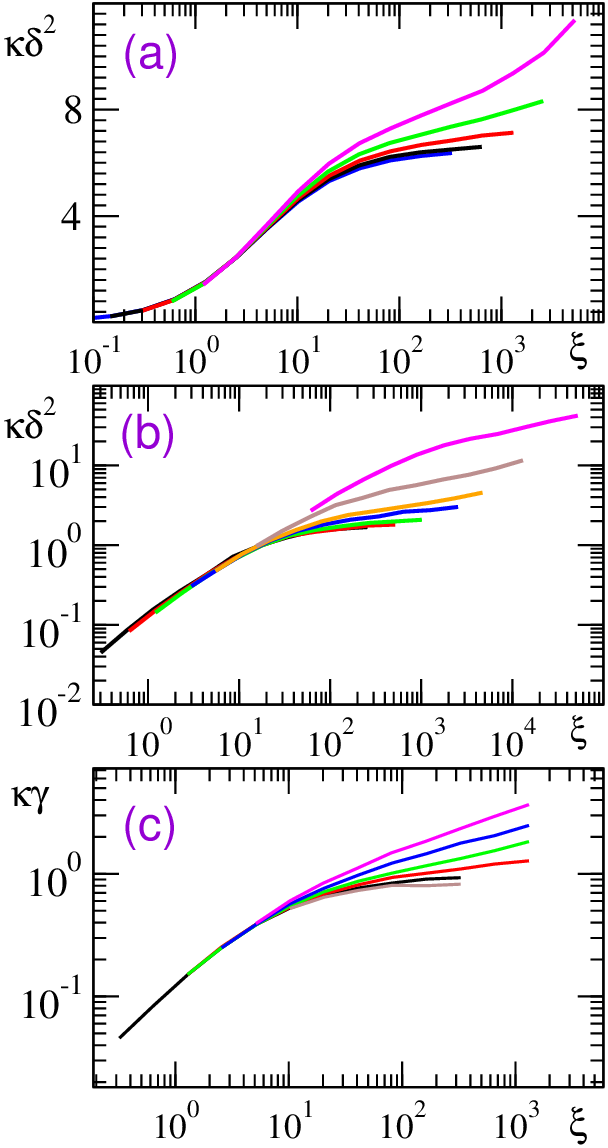}
%\end{center}
\caption{Nonequilibrum simulations: scaled thermal conductivity versus 
scaled chain lengths for (a) the HPG model $T_L=6,T_R=4$ $M=1$
$\delta=0.05,0.07,0.10,0.14,0.2$ from bottom to top;
(b) the diatomic Toda model (data taken from Ref.\cite{Chen2014}, Fig.4)
$\delta=0.07,0.10,0.14,0.22,0.30,0.50,1.0$ from bottom to top 
and (c) the Toda model with conserving noise 
$\gamma=0.005,0.01,0.02,0.04,0.08,0.16$ from bottom to top.
}
\label{fig:figure1}
\end{figure}

Here, we are interested in the quasi-integrable regime, $\delta,\gamma \ll 1$.
The first issue is the determination of the mean free path $\ell$. In the HPG
with a diatomic mass arrangement, $\ell$ corresponds to the average
space travelled by a single velociton before the collisions induce a sizeable change
of its original velocity.
Given the mass arrangement, the collision types $(m_1,m_2)$ and $(m_2,m_1)$ alternate
and it is thereby appropriate to look at velocity changes every second iterate
($u\to u'\to u''$).
From Eq.~(\ref{coll}) 
\begin{equation}
u''= (1\!-\!\delta)u'+ \delta \chi_2 = 
(1\!-\! \delta^2)u+\delta (\chi_2 -\chi_1) + \delta^2\chi_1 
\label{collmod}
\end{equation}
where $\chi_{1,2}$ denote the velocities of the quasi-particles encountered by the velociton $u$,
hereby assumed to be uncorrelated Gaussian variables with zero average and variance
$\langle \chi_{1,2}^2\rangle= v^2$ (neglecting the mass difference
between the two particles).
To leading order in $\delta$, the map can be turned into the stochastic differential equation
\[
\dot u =  -\delta^2 u + \sqrt{2}\delta v \zeta \; ,
\]
where $\zeta$ is a unit-variance white noise, while time is measured in $2\tau$ units, where
$\tau$ is the average collision time.
As a result, $u$ diffuses, its variance growing initially 
as $D_u =2\delta^2 v^2 \,t$, so that the time needed for $D_u$ to be approximately equal to $v^2$ is
$t \approx \tau/\delta^2$ (in physical time units) and the corresponding mean free path is 
$\ell \approx v \tau/\delta^2$. In other words, we expect $\theta=2$. 

Numerical results have been obtained by implementing the standard non-equilibrium procedure~\cite{LLP03,DHARREV}.
Left and right edges are attached to Maxwellian heat baths at temperatures $T_L=6$ and $T_R=4$ ($\Delta T = T_L-T_R$) 
and the flux $J$ determined from the average energy exchanged in the steady state. 
For the HPG we employ the thermal-wall method as detailed, e.g., in Ref.~\cite{Chen2014}. 

In Fig.~\ref{fig:figure1}(a) we plot the rescaled thermal conductivity 
$\delta^2 \kappa$ of the diatomic HPG, referred to the effective length 
$\xi=L\delta^2$~\footnote{Here and in the following, we take into account only the scaling of $\ell$ with 
$\varepsilon$, neglecting irrelevant multiplicative factors.}
(the various curves correspond to different $\delta$ values -- $\delta$ decreases from top to bottom).
There is clear evidence of a ballistic regime followed by a diffusive one,
as accounted for by Eq.~(\ref{JN}). For the smaller $\delta$'s, 
Fourier-like trasport persists up to the maximal available $L$, while 
a crossover to the anomalous regime is seen upon increasing $\delta$ (see the uppermost curve).
Upon decreasing $\delta$, the curves converge from above towards an asymptotic shape, $\kappa_N = j_N(\xi)\xi/\Delta T$.
In fact, for $\delta\to 0$ and fixed $\xi$, $L$ increases ($L=\xi/\delta^2$), so that
the corresponding $J_A(L)$ contribution becomes increasingly negligible.

\begin{figure}
%\begin{center}
\includegraphics[width=0.45\textwidth]{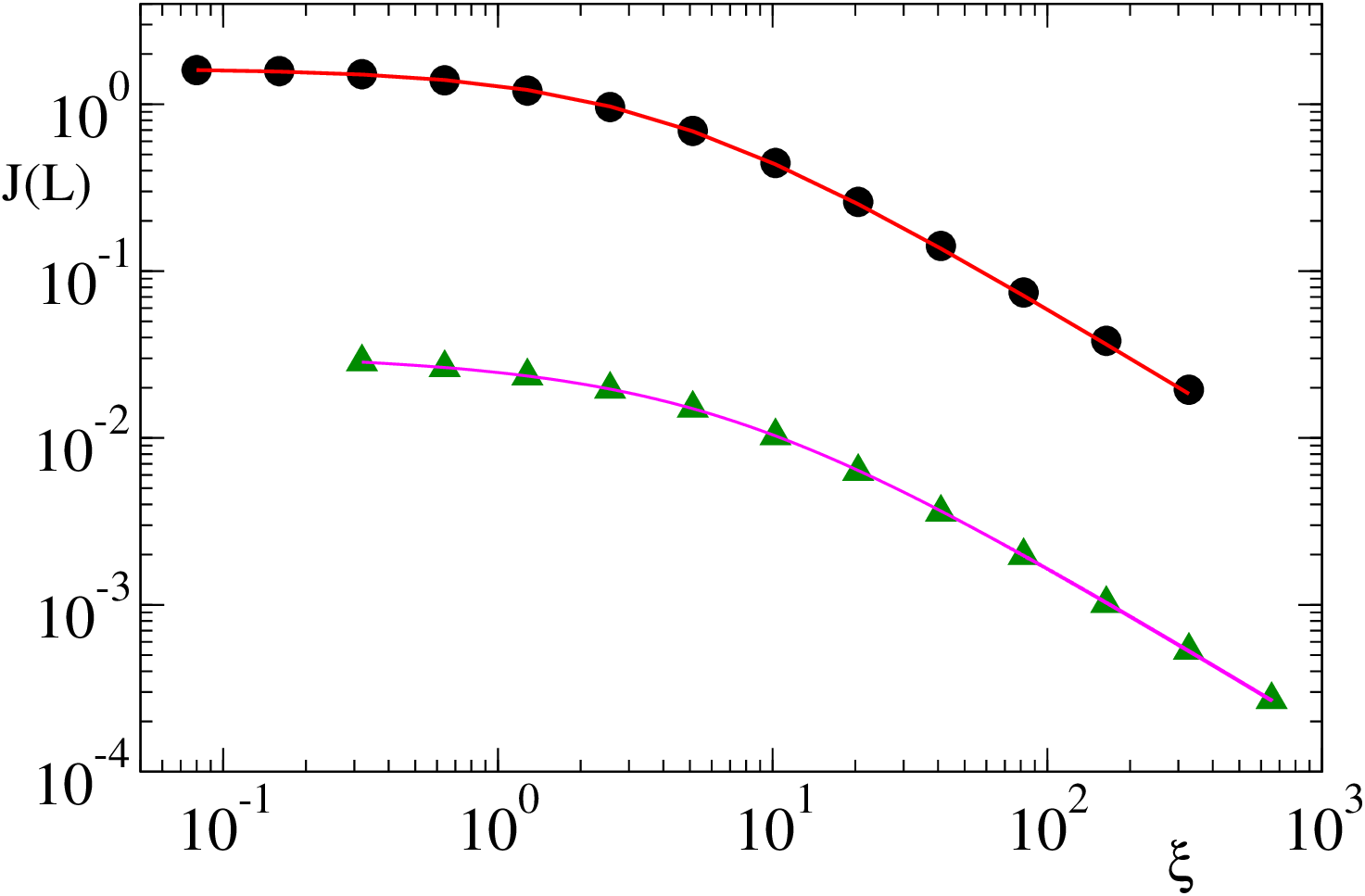}
%\end{center}
\caption{Estimates of dependence of the normal component $J_N(L)$ of the 
energy flux, as defined by Eq. (\ref{JN}), versus the rescaled 
system length $L$. 
Filled circles are for the HPG model with mass parameter $\delta=0.05$
and triangles for the Toda with random collisions $\gamma=0.005$.
The solid lines  are best fit with the functions $6.18/(3.73+\xi)$
and $1/(27.0+5.62\xi+7.87\xi^{0.2})$ respectively.
}
\label{fig:figure2}
\end{figure}

Then, $j_N$ is estimated from the data for the smallest
perturbation amplitude (since $J_A$ is practically negligible over the explored length range).
The simple formula~(\ref{kappad}) proves remarkably accurate:
see the solid upper curve in Fig.~\ref{fig:figure2}, to be compared with the circles, which represent the numerical HPG results.

As a second test, we consider the diatomic Toda model \citep{Chen2014}.
For large energy densities, most of its dynamical properties are basically equal
to the HPG \cite{Kundu2016} and we thus expect again $\ell \approx \delta^{-2}$, i.e. $\theta=2$. 
At lower energies, one should estimate the soliton scattering rates due to mass inhomogeneities; however
there is no reason to expect a different scaling behavior.
Indeed, the typical thermalization time is of order 
$\delta^{-2}$ in a wide energy range \cite{Fu2019}. 
The conductivities taken from Ref.~\cite{Chen2014}, are reported in Fig.~\ref{fig:figure1}(b) after a proper rescaling. 
The data collapse confirms the validity of our arguments.

Finally, we have considered the Toda model with conservative noise. In this case, it is natural to argue 
that the mean free path scales as the inverse of the collision rate, $\ell \sim \gamma^{-1}$.
This intuition is confirmed by the data reported in Fig.~\ref{fig:figure1}(c), where we observe the same scenario
as for the previous models, after setting $\theta =1$.
For small $\gamma$, $j_N$ converges again towards a function, which asymptotically decays as $1/\xi$. However, the 
heuristic formula~(\ref{kappad}) is not comparably accurate: it is necessary to add a correction term to reproduce
the observed data, as shown in Fig.~\ref{fig:figure2} (see triangles vs. the corresponding solid curve).

\begin{figure}
%\begin{center}
\includegraphics[width=0.45\textwidth]{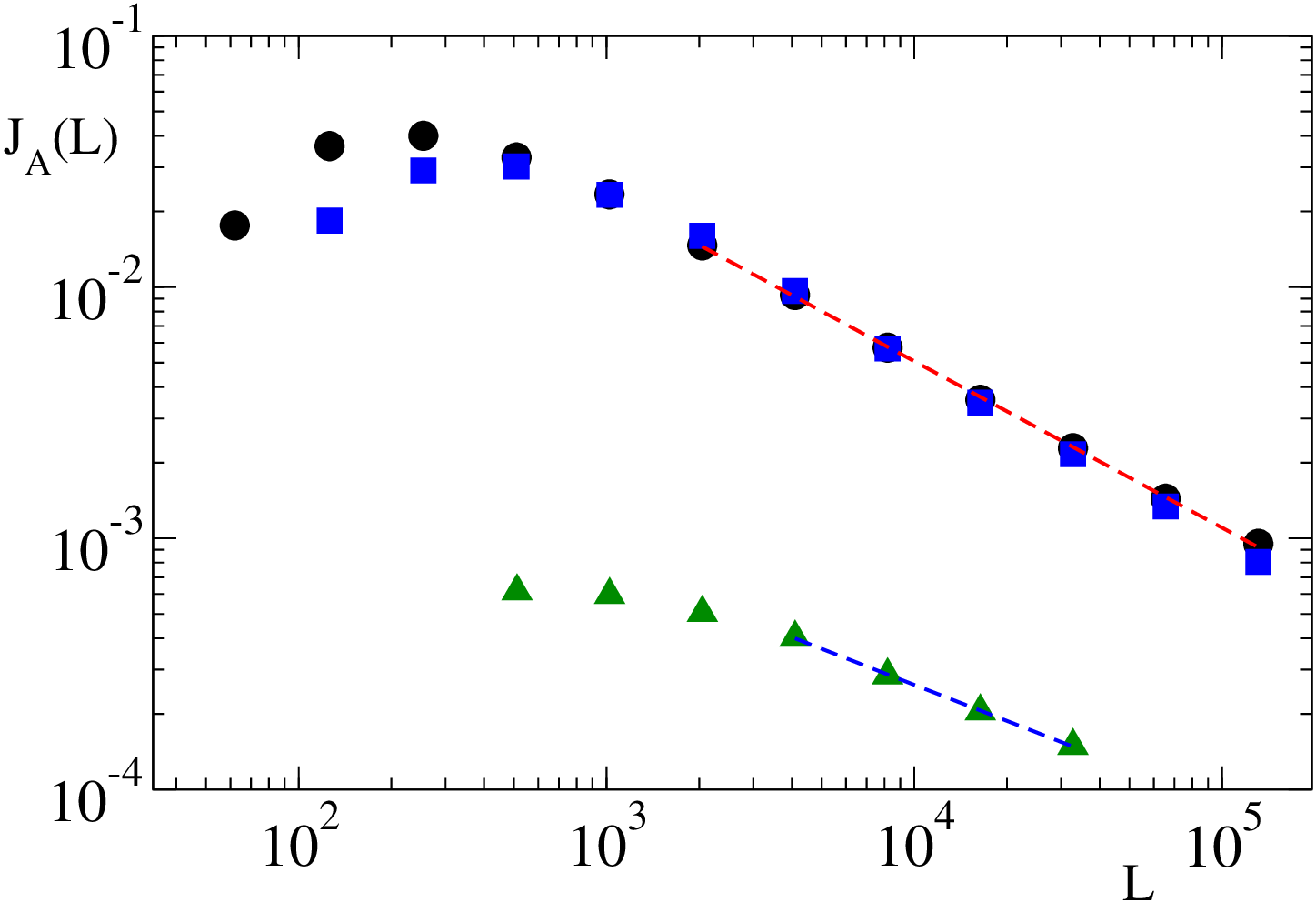}
%\end{center}
\caption{The anomalous part of the energy flux versus the system length $L$ calculated as $J_A = J(L)-J_N(L)$ with $J_N$ as determined in the previous figure. HPG model with $\delta=0.14$ (squares) and
$\delta=0.20$ (circles) and for Toda with conservative noise 
$\gamma=0.04$ (triangles). The dashed lines are power-law fit on the 
largest sizes where scaling sets in (values given in the text). 
}
\label{fig:figure3}
\end{figure}

So far we have tested the structure of the first addendum in Eq.~(\ref{Jtot}) by studying a regime
where the second contribution is negligible. What about the second addendum?
Once $j_N$ has been determined, one can proceed by estimating the anomalous component 
as $J_A(L) = J(L,\varepsilon) - j_N(L\varepsilon^\theta)$.
The data in Fig.~\ref{fig:figure3} indicates that $J_A$ exhibits the expected anomalous scaling already for system
sizes where the direct estimates are strongly affected by the diffusive component.
For the HPG the fitted slope, about $-0.66$, corresponds to $\eta=0.33$, in excellent agreement with the 
KPZ prediction $\eta=1/3$. For Toda we obtain $\eta=0.52$, consistent with Ref.~\cite{Iacobucci2010} and
even closer to $\eta=1/2$, the value rigorously proven for harmonic models with momentum-conserving noise 
\cite{BBO06,Basile08,Lepri2009}.
We thus conclude that the measurements confirm the above proposed crossover from diffusive to hydrodynamic 
behavior.

The same scenario is expected to emerge in the presence of a generic momentum-conserving perturbation
$\varepsilon W(y)$ of the potential of the Toda chain. 
In fact, in this case, it has been
already noticed that the energy-flux correlation decays over a time scale inversely proportional 
to a power of $\varepsilon$~\cite{Shastry2010}.

More in general, we argue that the crossover from
normal to anomalous regimes of the FPUT$-\alpha\beta$ model \cite{Wang2013,Das2014} is fully accounted 
for by the above described scenario. Indeed, the FPUT$-\alpha\beta$ 
(at low enough energies) can be regarded as a perturbed Toda chain over very long time scales,
on which the Toda actions are only weakly perturbed \cite{Benettin2013}. 
We reckon that other potentials should display the same
phenomenology, if their form is ``close enough" to $U_{T}$.

What can we say about the simplest model of a perturbed harmonic chain? 
This textbook case deserves a special consideration. 
Numerical analysis of the FPUT$-\beta$ model at very low energy,
i.e. below the strong stochasticity threshold, does not reveal any signature of
an intermediate diffusive regime, but rather a direct crossover 
from ballistic to anomalous regimes \cite{Lepri05}.
More compelling evidence of the absence of a diffusive regime comes from the study 
of the harmonic chain with conservative noise~\cite{BBO06} in the limit
of vanishing noise, i.e. $\gamma\to 0$. 
It has been found analytically~\cite{Lepri2009} and confirmed numerically~\cite{Delfini10} 
that $J_A(L,\varepsilon)$, exhibits a singular dependence for $\varepsilon \to 0$ (here $\gamma \to 0$)
in the form of a divergence of the coefficient $c_A$ in Eq.~(\ref{Jtot2}), $c_A \approx \gamma^{-1/2}$,
which implies that $J_A$ prevails over $J_N$ for any value of $L$.

Hence, the different behavior displayed by weakly perturbed harmonic oscillators can be traced back to a divergence
of the anomalous component, which is itself a consequence of the nonlinearity of the dynamical equations.
 This counter-intuitive phenomenon reminds us that Eq.~(\ref{Jtot}) is a conjecture, which still requires
a rigorous derivation. Nevertheless, the successfull implementation of our scaling arguments provides 
a convincing explanation of
the seemingly normal diffusion observed not only in Refs.~\cite{Iacobucci2010,Chen2014,Zhao2018}, used as our testbeds,
but also in many  nonlinear chains like those discussed in Refs.~\cite{Wang2013,Das2014} 
where the potential is ``well approximated" by the Toda one.

\acknowledgments
SL and RL acknowledge partial support from project MIUR-PRIN2017 \textit{Coarse-grained description for non-equilibrium systems and transport phenomena (CO-NEST)} n. 201798CZL 
%and the Eurofusion Enabling Research ENR-MFE19.CEA-06 \textit{Emissive divertor}. 
SL is grateful to S. Ruffo and 
the program 
\textit{Collaborazioni di Eccellenza} of SISSA,
Trieste, Italy.

\end{document}